\documentclass[%
reprint, 
showkeys,
amsmath,amssymb,
aps,pra,
floatfix,
]{revtex4-2}

\usepackage{graphicx,graphics,times,bm,bbm,bbold,mathrsfs,amsfonts,dsfont,color}
\usepackage{setspace}
\usepackage{soul,xcolor,verbatim}
\setstcolor{red}
\usepackage{hyperref}
\hypersetup{colorlinks,linkcolor=blue,citecolor=blue,urlcolor=blue}
\usepackage{dcolumn}

\usepackage{tikz}
\usepackage{tikz-qtree}
\usepackage{yquant}
\useyquantlanguage{qasm}

\newcommand{\ket}[1]{\big| #1 \big\rangle}
\newcommand{\bra}[1]{\big\langle #1 \big|}
\newcommand{\braket}[2]{\big\langle #1 \big| #2 \big\rangle}                 

\begin{document}

\title{Implementation of Continuous-Time Quantum Walks on Quantum Computers}
\author{Renato Portugal and Jalil Khatibi Moqadam}
\affiliation{%
 National Laboratory of Scientific Computing (LNCC)\\
 Petr\'opolis, RJ, 25651-075, Brazil 
}%


\date{\today}

\begin{abstract}
Quantum walk is a useful model to simulate complex quantum systems and to build quantum algorithms; in particular, to develop spatial search algorithms on graphs, which aim to find a marked vertex as quickly as possible. Quantum walks are interesting candidates to be implemented on quantum computers. In this work, we describe efficient circuits that implement the evolution operator of continuous-time quantum-walk-based search algorithms on three graph classes: complete graphs, complete bipartite graphs, and hypercubes. For the class of complete and complete bipartite graphs, the circuits implement the evolution operator exactly. For the class of hypercubes, the circuit implements an approximate evolution operator, which tends to the exact evolution operator when the number of vertices is large. Our Qiskit simulations show that the implementation is successful at finding the marked vertex even for low-dimensional hypercubes.
\end{abstract}



\maketitle

\section{Introduction}
\label{sec:intro}


The continuous-time quantum walk (CTQW) was introduced by Farhi and Gutmann~\cite{FG98} as the quantum version of the continuous-time Markov chain, and this new model proved useful to build quantum algorithms, such as algorithms for spatial search on graphs~\cite{CG_04, ABM_10,PTB_16,OCOSMN_20} and NAND-based formula evaluation~\cite{FGG_08}. Experimental implementations of search algorithms by continuous-time quantum walk are described in~\cite{DGGCWS_19,DGPSW20,WSXWJX20,BTPC21,QMWXWX22}.

Circuits for the implementation of CTQWs on graphs with no marked vertex are obtained using Hamiltonian simulation, for instance, Refs.~\cite{QLMAZOWM16,SCC21} describe circuits for the CTQW on circulant graphs using this method. Hamiltonian simulation is different from the methods used to implement discrete-time quantum walks~\cite{ADZ93,Por16b}, such as the ones described in~\cite{AAMP20,NZDPDA22,GEZ21}.

In this work, we tackle the problem of finding efficient circuits for the implementation of CTQW-based search algorithms on graphs with one marked vertex. We focus on three graph classes: complete graphs, complete bipartite graphs, and hypercubes. The techniques described here can be applied to other graph classes. When the graph structure is simple enough, such as complete graphs and complete bipartite graphs, we build a circuit that simulates the exact evolution operator $U$, modulo a global phase. On the other hand, for hypercubes, we build a circuit that implements a unitary operator $U'$ that simulates $U$ approximately, not in the sense that the Hilbert-Schmidt distance between $U$ and $U'$ is small, but in the sense that the difference between the probability of finding the marked vertex using $U$ and $U'$ is small when the initial condition is the uniform superposition $\ket{s}$ of all states of the computational basis.

Our method relies on the fact that, for some graph classes, the fidelity between the uniform state $\ket{s}$ with the plane spanned by the ground state $\ket{\lambda^-}$ and first excited state $\ket{\lambda^+}$ of the Hamiltonian tends to 1 asymptotically ($N\rightarrow\infty$, where $N$ is the number of vertices). For these graph classes, we describe a method to calculate approximations of $\ket{\lambda^\pm}$. Since the Hamiltonian is described only in terms of projectors on the eigenspaces of these two eigenvectors, we use the fact that they commute, and then the evolution operator can be written as a product of two unitary operators $U=U_{\lambda^+}U_{\lambda^-}$. In the final step, we implement $U_{\lambda^+}$ and $U_{\lambda^-}$ using circuits for state preparation, that is, using the circuit of an operator $A_{\lambda}$ such that $A_{\lambda}\ket{0}=\ket{\lambda}$, where $\ket{0}$ is the first state of the computational basis.

In all graph classes that we have addressed, we have obtained circuits for the evolution operators with $O(\log^2 N)$ basic gates. However, to run the whole search algorithm we need $O(\sqrt N)$ steps, that is, at the end, we implement $U^{\left\lfloor t_\text{opt}\right\rfloor}$, where $t_\text{opt}$ is the optimal running time of the algorithm. Therefore, the circuits have $\tilde O(\sqrt N)$ basic gates.

The structure of the paper is as follows. Sec.~\ref{sec:2} reviews the continuous-time quantum walk dynamics on graphs. The following Sections present the decomposition of the evolution operator and the associated circuit for two kinds of CTQW on graphs, namely, graphs with no marked vertex, and graphs with one marked vertex. We have addressed three graph classes:~(1) complete graphs in Sec.~\ref{sec:complete_graph},~(2) complete bipartite graphs in Sec.~\ref{sec:complete_bipartite_graph}, and~(3) hypercubes in Sec.~\ref{sec:hypercube}. In Sec.~\ref{sec:conclusion}, we present our final remarks.

\section{CTQW on graphs}
\label{sec:2}
A continuous-time quantum walk~\cite{FG98} on a graph $\Gamma(V,E)$, with vertex set $V$ and edge set $E$, is a quantum dynamics in which the state space is associated with $V$, and the evolution operator is $U(t)=\text{e}^{-i {\mathcal{H}} t}$, where ${\mathcal{H}}=-\gamma A$ is the Hamiltonian and $A$ the adjacency matrix, which is a symmetric matrix whose entries $A_{k\ell}$ are 1 for all pairs of vertices $(v_k,v_{\ell})\in E$, and 0 otherwise. The hopping probabilities per unit time between neighboring vertices are given by a positive transition rate $\gamma$. If the initial state is $\ket{\psi(0)}$, the state of the quantum walk at time $t$ is $\ket{\psi(t)}=U(t)\ket{\psi(0)}$.

\subsection{Spatial search on graphs}
The CTQW is an interesting framework to develop spatial search algorithms on graphs. These algorithms aim to find a marked vertex $w\in V$ as quickly as possible starting from an initial state uniformly spread over all vertices. The standard recipe is driven by the modified Hamiltonian~\cite{CG_04}
\begin{equation}
    \label{eq:oracle_Hamiltonian}
    {\mathcal{H}} = -\gamma A - | w \rangle\langle w |.
\end{equation}
The CTQW-based search provides a quadratic speedup when compared with a random-walk-based search on the complete graph~\cite{CG_04}, bipartite graph, hypecube~\cite{CG_04}, and Johnson graph~\cite{TSP22}, to mention a few of them. However, the CTQW-based search algorithms do not outperform the classical algorithms on $d$-dimensional lattices with $d<4$~\cite{CG_04}. 

The efficiency of the algorithm is determined by the optimal running time $t_\text{opt}$ so that the success probability $p_\text{succ}=\left|\langle w | \psi(t_\text{opt}) \rangle\right|^2$ is as large as possible, where 
\begin{equation}\label{eq:psi-t}
\ket{\psi(t)}=\left(\text{e}^{-i {\mathcal{H}} }\right)^{t}\ket{\psi(0)},
\end{equation}
in which ${\mathcal{H}}$ is the modified Hamiltonian (\ref{eq:oracle_Hamiltonian}), and $\ket{\psi(0)}$ is the initial condition
\begin{equation}\label{eq:initial_condiction}
| \psi(0) \rangle = \ket{s} = \frac{1}{\sqrt{N}}\sum_{j=0}^{N-1} |j\rangle,
\end{equation}
where $N$ is the number of vertices.

\subsection{Locality}\label{subsec:loc}

Quantum walk is a subarea of quantum mechanics that is characterized essentially by two aspects: (1) the allowed locations of the walker is a spatial discrete structure, usually modeled by a graph, and (2) the evolution operator is local in the sense that if the walker is on vertex $v$, the walker hops to the neighboring vertices of $v$ before reaching the other vertices. The dynamics~\eqref{eq:psi-t} satisfies the two aspects by the following reasons: Let the initial condition be $\ket{\psi(0)}=\ket{v}$ and take $\gamma$ constant for all graphs in a graph class. After an infinitesimal time $\epsilon$, the locations of the walker is the superposition state
\begin{equation}
\ket{v} + \epsilon\left(\gamma A+\delta_{v,w}\right)\ket{v} + O\left(\epsilon^2\right)\cdots \,+,
\end{equation}
which satisfies the locality aspect because the action of the adjacency matrix on $\ket{v}$ results in a superposition of vertices in the neighborhood of $v$.

If we want to compare the time complexity of CTQW-based with random-walk-based search algorithms, we have to change the viewpoint. Since the random walk evolves in discrete time-steps, a fair comparison demands that we take $t=1$ in~\eqref{eq:psi-t} and repeat the action of $\text{e}^{-i {\mathcal{H}} }$ over and over. To satisfy the locality aspect in this case, we take $\gamma$ small, typically $O\left(1/N\right)$, where $N$ is the number of vertices. The optimal value of $\gamma$ of the CTQW-based search algorithm on the class of complete graphs and complete bipartite graphs is $O\left(1/N\right)$. On the other hand, the optimal value of $\gamma$ for the class of hypercube graphs is $O\left(1/\log N\right)$, which satisfies a weaker version of the locality aspect.

\section{\label{sec:complete_graph}Complete graph}

\begin{figure}[!t]
\begin{tikzpicture}
\begin{yquant}
qubit {} q[4];
box {$H$} q[0-3];
box {$X$} q[0-3];
box {$\text{e}^{\frac{it}{2}}R_z({t})$} q[3] | q[0-2];
box {$X$} q[0-3];
box {$H$} q[0-3];
\end{yquant}
\end{tikzpicture}
\caption{\label{fig:complete-graph-1}Circuit implementing the CTQW on the complete graph $K_{N}$, where $U(t)$ is given by Eq.~(\ref{eq:U-complete-graph}) and $N=16$.}
\end{figure}
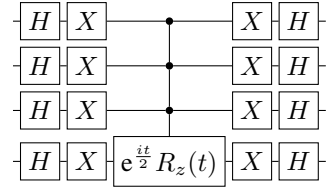

Let $K_{N}$ be the complete graph, where $N=2^n$. The adjacency matrix of $K_N$ is
\begin{equation}
    A=N |s \rangle\langle s|-I,
\end{equation}
where $| s \rangle$ is the uniform superposition of all vertices or all basis states, $| s \rangle=H^{\otimes n}| 0 \rangle,$
and $H=\scriptstyle \frac{1}{\sqrt{2}}\big(\begin{smallmatrix}1&\hspace{0.2cm}1\\1&-1\end{smallmatrix}\big)$ is the Hadamard operator. Assuming that $\gamma=1/N$, the evolution operator of a CTQW on $K_N$ with no marked vertex is
\begin{equation}
    U(t)=\text{e}^{\frac{-it}{N}}H^{\otimes n}\text{e}^{it|0 \rangle\langle 0|}H^{\otimes n}.
\end{equation}
Using the identity
\begin{equation}\label{eq:Identity_P}
    \text{e}^{itP}=I+(\text{e}^{it}-1)P,
\end{equation}
which is true for any orthogonal projector $P$, we obtain
\begin{equation}\label{eq:U-complete-graph}
    U(t)=\text{e}^{\frac{-it}{N}}H^{\otimes n}\big(I+(\text{e}^{it}-1)|0 \rangle\langle 0|\big)H^{\otimes n}.
\end{equation}
The circuit that implements $U(t)$ up to a global phase is depicted in Fig.~\ref{fig:complete-graph-1} for the case $n=4$. The implementation of 
\[
R=I+(\text{e}^{it}-1)|0 \rangle\langle 0|
\]
uses a multi-controled gate 
\[
R_z(\theta)=\scriptstyle \left(\begin{matrix}\text{e}^{-i\theta/2}&\hspace{0.2cm}0\\0&\text{e}^{i\theta/2}\end{matrix}\right)
\]
multiplied by a relative phase $\text{e}^{\frac{i \theta}{2}}$. Note that $R\ket{0}=\text{e}^{it}\ket{0}$ and $R\ket{j}=\ket{j}$ if $j\neq 0$. If $n=1$, $R$ is obtained as follows
\begin{equation}
   R=\text{e}^{\frac{i t}{2}} X R_z(t)X=
\left(\begin{array}{cc}
\text{e}^{ i t}  &  0 \\
 0   &  1
\end{array}\right).
\end{equation}
If $n>1$, we have to use $n-1$ control qubits that are activated only when they are set to 0.

In Qiskit~\footnote{\url{https://qiskit.org/}}, the multi-controled gate $\text{e}^{\frac{i t}{2}} R_z(t)$ is implemented using gate \verb|mcrz|~\footnote{See package \textit{qiskit.circuit.QuantumCircuit}.}, and its decomposition in terms of universal gates is automatically provided.

\subsection{Spatial search on complete graph}


We take $\gamma=1/N$ because this is the asymptotic optimal value for the spatial search algorithm on the complete graph $K_N$~\cite{CG_04}. Using this value in the modified Hamiltonian (\ref{eq:oracle_Hamiltonian}), we obtain
\begin{equation}
{\mathcal{H}}=\frac{I}{N}-|s \rangle\langle s|-|w \rangle\langle w|
\end{equation}
where $w$ is the marked vertex. Alternatively, we write
\begin{equation}
{\mathcal{H}}-\frac{I}{N}=\lambda^+|\lambda^+ \rangle\langle \lambda^+| +\lambda^-|\lambda^- \rangle\langle \lambda^-|,
\end{equation}
where $|\lambda^\pm \rangle$ are the only eigenvectors of ${\mathcal{H}}-I/N$ associated with non-zero eigenvalues $\lambda^\pm$, which are given by
\begin{equation}
\lambda^\pm = -1 \pm \frac{1}{\sqrt N}
\end{equation}
and the entries of the eigenvectors are
\begin{equation}
\braket{j}{\lambda^\pm}= \begin{cases}
     \frac{\sqrt{- \lambda^\pm}}{\sqrt{2 }} & \text{if }j=w, \\
   \frac{ \mp 1}{\sqrt{-2 N \lambda^\pm}} &  \text{if }j\neq w.
\end{cases}
\end{equation} 
In the continuation, we assume that $w=0$. The eigenvectors in this case are
\begin{equation}
\ket{\lambda^\pm} = \frac{1}{\sqrt{-2 N \lambda^\pm}}
\begin{pmatrix}
  -\sqrt{N}\lambda^\pm \\
  \mp 1 \\
  \vdots \\
  \mp 1 \\
\end{pmatrix}.
\end{equation}   

Using that the projectors $|\lambda^+ \rangle\langle \lambda^+|$ and $|\lambda^- \rangle\langle \lambda^-|$ commute, the evolution operator of a CTQW on $K_N$ with one marked vertex is
\begin{equation}
U(t)= \text{e}^{-it/N}\text{e}^{-it \lambda^+|\lambda^+ \rangle\langle \lambda^+|}\text{e}^{-it\lambda^-|\lambda^- \rangle\langle \lambda^-|}.
\end{equation}
Using Eq.~(\ref{eq:Identity_P}), we obtain
\begin{equation}\label{eq:U-complete-graph-search}
U(t)= \text{e}^{-it/N} A_{ \lambda^+}R_{\lambda^+}(t)A_{\lambda^+}^\dagger A_{\lambda^-}R_{\lambda^-}(t)A_{\lambda^-}^\dagger
\end{equation}
where
\begin{equation}
R_{\lambda^\pm}(t) = I+(\text{e}^{-it\lambda^\pm}-1)|0\rangle\langle 0|
\end{equation}
and~\footnote{The choice of $\ket{0}$ in this equation has no relation with the location of the marked vertex.}
\begin{equation}\label{eq:A-complete-graph}
A_{\lambda^\pm}\ket{0}=|\lambda^\pm \rangle.
\end{equation}

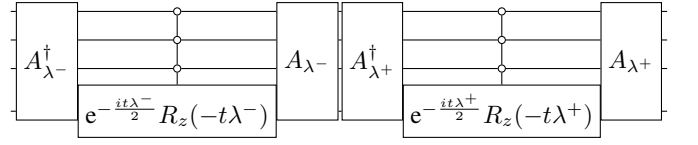
\begin{figure}[!ht]
\begin{tikzpicture}[scale=0.95]
\begin{yquant}
qubit {} q[4];
box {$A_{\lambda^-}^\dagger$} (q[0-3]) ;
box {$\text{e}^{-\frac{it\lambda^-}{2}}R_z(-t\lambda^-)$} q[3] ~ q[2-0];
box {$A_{\lambda^-}$} (q[0-3]) ;
box {$A_{\lambda^+}^\dagger$} (q[0-3]) ;
box {$\text{e}^{-\frac{it\lambda^+}{2}}R_z(-t\lambda^+)$} q[3] ~ q[2-0];
box {$A_{\lambda^+}$} (q[0-3]) ;
\end{yquant}
\end{tikzpicture}
\caption{\label{fig:U-complete-graph-search}The circuit that implements $U(t)$~(\ref{eq:U-complete-graph-search}) on $K_{2^n}$ with one marked vertex when $n=4$.  The circuit of $A_{\lambda^\pm}$ is depicted in Fig.\ref{fig:A-complete-graph};}
\end{figure}

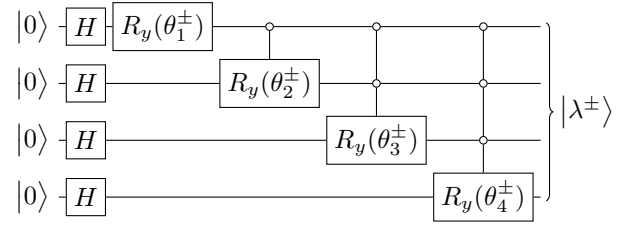
\begin{figure}[!ht]
\begin{tikzpicture}
\begin{yquant*}
qubit {$\ket{0}$} q[4];
box {$H$} q[0-3];
box {$R_y(\theta^\pm_1)$} q[0];
box {$R_y(\theta^\pm_2)$} q[1] ~ q[0];
box {$R_y(\theta^\pm_3)$} q[2] ~ q[1-0];
box {$R_y(\theta^\pm_4)$} q[3] ~ q[2-0];
output {$\ket{\lambda^\pm}$} (q[0-3]);
\end{yquant*}
\end{tikzpicture}
\caption{\label{fig:A-complete-graph}The circuit that implements $A_{\lambda^\pm}$~(\ref{eq:A-complete-graph}) when $n=4$. Angles $\theta_k^\pm$ are given by Eq.~(\ref{eq:theta-complete-graph-1}). }
\end{figure}

The circuit that implements $U(t)$~(\ref{eq:U-complete-graph-search}) is depicted in Fig.~\ref{fig:U-complete-graph-search}. The circuit of $A_{\lambda^\pm}$ is obtained using the techniques described in Appendix~\ref{appendix:A} and is depicted in Fig.~\ref{fig:A-complete-graph}. Using Eqs.~(\ref{eq:a-k-appen}), (\ref{eq:theta-k-prime}), and~(\ref{eq:theta-k}), the angles of operators $R_y$ when $w=0$ are 
\begin{equation}\label{eq:theta-complete-graph-1}
\theta^\pm_k =  \mp 2\arctan \frac{1}{\sqrt{1+2^{k}\mp \frac{2^{k+1}}{\sqrt N}}}-\frac{\pi}{2},
\end{equation}
for $1\le k \le n$.

If we consider a finite set of basic gates, the number of basic gates required to implement the $A_{\lambda^\pm}$ in Fig.~\ref{fig:A-complete-graph} is $O(n^2)$ for the angles of Eq.~(\ref{eq:theta-complete-graph-1}).
The computational cost to implement $R_{\lambda^\pm}(t)$ depends on $t$. For a fixed $t$, let's say $t=1$, the number of universal gates is also $O(n^2)$ because $\lambda^\pm\approx -1$ for large $N$.
For the search algorithm, the optimal running time is~\cite{CG_04} 
\begin{equation}
t_\text{opt}= \frac{\pi}{2}\sqrt{N},
\end{equation}  
which means that the success probability is exactly 1 at $t_\text{opt}$. In terms of oracle-based algorithms, the oracle is the circuit of $\text{e}^{-i{\mathcal{H}}}$, obtained by taking $t=1$ in the circuit of Fig.~\ref{fig:U-complete-graph-search}. To find the marked vertex, in this case $w=0$, one must concatenate the circuit of Fig.~\ref{fig:U-complete-graph-search} $\lfloor t_\text{opt}\rfloor$ times. Therefore, the implementation requires $\tilde{O}(\sqrt{N})$ basic gates. Note that operator $\text{e}^{-i{\mathcal{H}}}$ is local in the sense that the walker hops only to neighboring vertices after the action of $\text{e}^{-i{\mathcal{H}}}$ because the hopping rate is $\gamma=1/N$. To compare the continuous-time evolution with a discrete-time random-walk-based evolution, we have to repeat the action of $\text{e}^{-i{\mathcal{H}}}$ over and over instead of setting $t=\pi\sqrt{N}/2$ in the circuit of Fig.~\ref{fig:U-complete-graph-search} (see discussion in Subsec.~\ref{subsec:loc}).

\begin{figure}[!ht]
\includegraphics[scale=0.55]{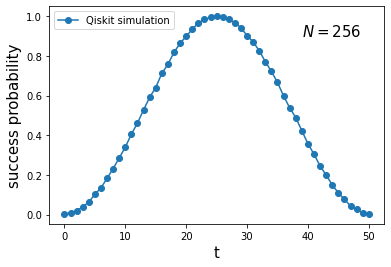}
\caption{Success probability $\left|\braket{0}{\psi(t)}\right|^2$ as a function of the number of steps $t$ for the search on $K_{2^n}$ with $n=8$ using a Qiskit implementation of the circuit of Fig.~\ref{fig:U-complete-graph-search}.}\label{fig:complete-n-8}
\end{figure}

Fig.~\ref{fig:complete-n-8} depicts the success probability $\left|\braket{0}{\psi(t)}\right|^2$ as a function of the number of steps $t$ for a complete graph  $K_{256}$. The curve is obtained using the implementation of the circuit of Fig.~\ref{fig:U-complete-graph-search} in Qiskit. Since $t$ is discrete, the optimal running time is $\left\lfloor t_\text{opt}\right\rfloor$.

 


\section{Complete bipartite graph}\label{sec:complete_bipartite_graph}

Let $K_{n,n}$ be the complete bipartite graph with $N=2n=2^m$ vertices and adjacency matrix $A$. Define ${\tilde A}$ as
\begin{equation}
    {\tilde A}=H^{\otimes m} AH^{\otimes m},
\end{equation}
which is a remarkably simple matrix given by
\begin{equation}
{\tilde A}=n\,|0 \rangle\langle 0|-n\,|n \rangle\langle n|,
\end{equation}
whose nonzero eigenvalues are $\lambda^\pm=\pm n$ with associated eingenvectors $\ket{\lambda^+}=\ket{0}$ and $\ket{\lambda^-}=\ket{n}$. Taking $\gamma=1/n$ and using that the projectors $|0 \rangle\langle 0|$ and $|n \rangle\langle n|$ commute, the evolution operator of a CTQW on the complete bipartite graph with no marked vertex is
\begin{align}
U(t)&=\text{e}^{i \gamma A t}=H^{\otimes m}\text{e}^{i \gamma {\tilde A} t}H^{\otimes m}\nonumber\\
&=H^{\otimes m} \text{e}^{i t|0 \rangle\langle 0|} \text{e}^{-i t|n \rangle\langle n|} H^{\otimes m}.
\end{align}
Using Eq.~(\ref{eq:Identity_P}), we obtain
\begin{equation}\label{eq:U-bipartite-graph}
U(t)=H^{\otimes m}R_0 R_n H^{\otimes m},
\end{equation}
where\vspace{-3pt}
\begin{align*}
R_0 &= I+ (\text{e}^{i t}-1)|0 \rangle\langle 0|,\\
R_n &= I+ (\text{e}^{-i t}-1)|n \rangle\langle n|.
\end{align*}

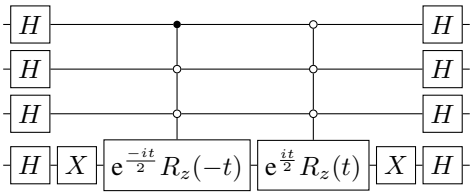
\begin{figure}[!ht]
\begin{tikzpicture}
\begin{yquant}
qubit {} q[4];
box {$H$} q[0-3];
box {$X$} q[3];
box {$\text{e}^{\frac{-it}{2}}R_z(-{t})$} q[3] | q[0] ~ q[1-2];
box {$\text{e}^{\frac{it}{2}}R_z({t})$} q[3] ~ q[0-2];
box {$X$} q[3];
box {$H$} q[0-3];
\end{yquant}
\end{tikzpicture}
\caption{\label{fig:bipartite-graph-1}Circuit implementing the CTQW on the bipartite graph $K_{n,n}$ with $n=8$, where $U(t)$ is given by Eq.~(\ref{eq:U-bipartite-graph}).}
\end{figure}

The circuit that implements $U(t)$ up to a global phase is depicted in Fig.~\ref{fig:bipartite-graph-1} for the case $m=4$. Note that the first control qubit of $R_n$ is solid because $n=(10...0)_2$.

\subsection{Spatial search on complete bipartite graph}

\begin{figure*}[!ht]
\begin{tikzpicture}
\begin{yquant}
qubit {} q[4];
box {$A_{\lambda^-}^\dagger$} (q[0-3]) ;
box {$\text{e}^{-\frac{it\lambda^-}{2}}R_z(-t\lambda^-)$} q[3] ~ q[2-0];
box {$A_{\lambda^-}$} (q[0-3]) ;
box {$A_{\lambda^+}^\dagger$} (q[0-3]) ;
box {$\text{e}^{-\frac{it\lambda^+}{2}}R_z(-t\lambda^+)$} q[3] ~ q[2-0];
box {$A_{\lambda^+}$} (q[0-3]) ;
box {$A_{\lambda_0}^\dagger$} (q[0-3]) ;
box {$\text{e}^{-\frac{it\lambda_0}{2}}R_z(-t\lambda_0)$} q[3] ~ q[2-0];
box {$A_{\lambda_0}$} (q[0-3]) ;
\end{yquant}
\end{tikzpicture}
\caption{\label{fig:U-bipartite-graph-search}The circuit that implements $U(t)$~(\ref{eq:U-bipartite-graph-search}) on $K_{n,n}$ with one marked vertex when $n=8$.  The circuit of $A_{\lambda}$ is depicted in Fig.~\ref{fig:A-bipartite-graph};}
\end{figure*}


We take $\gamma=1/n$ because this is the asymptotic optimal value for the spatial search algorithm on the complete bipartite graph $K_{n,n}$, as can be obtained using the method described in Appendix~\ref{appendix:B}. Using $\gamma=1/n$ in the modified Hamiltonian (\ref{eq:oracle_Hamiltonian}), we obtain
\begin{equation}
{\mathcal{H}}=-H^{\otimes m}\left( |0\rangle\langle 0|-|n\rangle\langle n|\right)H^{\otimes m}-|w \rangle\langle w|
\end{equation}
where $w$ is the marked vertex. Since this Hamiltonian has only three eigenvalues different from 0, we write
\begin{equation}\label{eq:H-bipartite-lambda}
{\mathcal{H}}=\lambda^+|\lambda^+ \rangle\langle \lambda^+| +\lambda^-|\lambda^- \rangle\langle \lambda^-|+\lambda_0|\lambda_0 \rangle\langle \lambda_0|,
\end{equation}
where $\lambda^-<\lambda^+<\lambda_0$ are solutions of
\begin{equation}\label{eq:caracpol}
\lambda^3 + \lambda^2 - \lambda - \left(1-\frac{1}{n}\right) =0.
\end{equation}
When $w=0$, the entries $\lambda(j)=\braket{j}{\lambda}$ of associated eigenvectors are
\begin{equation}\label{eq:lambda-bipartite}
\lambda(j) = \begin{cases}
   \frac{1}{\sqrt{c}} &  \text{if } j=0, \\
   \frac{a}{\sqrt{c}} &  \text{if } 1\le j\le n, \\
   \frac{b}{\sqrt{c}} &  \text{if } n< j< N, 
\end{cases}
\end{equation}
where
\begin{align}
a &= \frac{n\lambda^2+n\lambda-1}{n-1},\\
b &= -1-\lambda,\\
c &= 1+(n-1)a^2+nb^2. 
\end{align}

Using that the projectors $|\lambda^+ \rangle\langle \lambda^+|$, $|\lambda^- \rangle\langle \lambda^-|$, and $|\lambda_0 \rangle\langle \lambda_0|$ commute, the evolution operator of a CTQW on $K_{n,n}$ with one marked vertex is
\begin{equation}
U(t)= \text{e}^{-it \lambda_0|\lambda_0 \rangle\langle \lambda_0|}\text{e}^{-it \lambda^+|\lambda^+ \rangle\langle \lambda^+|}\text{e}^{-it\lambda^-|\lambda^- \rangle\langle \lambda^-|}.
\end{equation}
Using Eq.~(\ref{eq:Identity_P}), we obtain
\begin{equation}\label{eq:U-bipartite-graph-search}
U(t)= A_{ \lambda_0}R_{\lambda_0}A_{\lambda_0}^\dagger  A_{ \lambda^+}R_{\lambda^+}A_{\lambda^+}^\dagger A_{\lambda^-}R_{\lambda^-}A_{\lambda^-}^\dagger
\end{equation}
where
\begin{equation}
R_{\lambda} = I+(\text{e}^{-it\lambda}-1)|0\rangle\langle 0|
\end{equation}
and
\begin{equation}\label{eq:A-bipartite-graph}
A_{\lambda}\ket{0}=|\lambda \rangle,
\end{equation}
where $\lambda$ is a root of~(\ref{eq:caracpol}) and $\ket{\lambda}$ is the associated eigenvector.

\begin{figure}[!ht]
\begin{tikzpicture}
\begin{yquant*}
qubit {$\ket{0}$} q[4];
box {$H$} q[0-3];
box {$R_y(\theta^\lambda_1)$} q[0];
box {$R_y(\theta^\lambda_2)$} q[1] ~ q[0];
box {$R_y(\theta^\lambda_3)$} q[2] ~ q[1-0];
box {$R_y(\theta^\lambda_4)$} q[3] ~ q[2-0];
output {$\ket{\lambda}$} (q[0-3]);
\end{yquant*}
\end{tikzpicture}
\caption{\label{fig:A-bipartite-graph}The circuit that implements $A_{\lambda}$~(\ref{eq:A-bipartite-graph}) when $m=4$. Angles $\theta_k^\lambda$ are given by Eq.~(\ref{eq:theta-bipartite-graph-1}). }
\end{figure}

The circuit that implements $U(t)$~(\ref{eq:U-bipartite-graph-search}) is depicted in Fig.~\ref{fig:U-bipartite-graph-search}. The circuit that implements $A_{\lambda}$ when $\ket{\lambda}$ is given by~(\ref{eq:lambda-bipartite}) is obtained using the techniques described in Appendix~\ref{appendix:A} and is depicted in Fig.~\ref{fig:A-bipartite-graph}. Using Eqs.~(\ref{eq:a-k-appen}), (\ref{eq:theta-k-prime}), and~(\ref{eq:theta-k}), the angles of operators $R_y$ when $w=0$ are
\begin{align}\label{eq:theta-bipartite-graph-1}
\theta^\lambda_k = \, & 2\arctan \frac{a\left(1-\delta_{km}\right)+b\delta_{km}}{\sqrt{a^2(1-2^{k-1})-b^2 2^{k-1}+c 2^{k-m}}}
 -\frac{\pi}{2},
\end{align}
for $1 \le k \le n$. Note that $a$, $b$, and $c$ depend on $\lambda$, which assumes the three roots of Eq.~(\ref{eq:caracpol}).


If we remove the gates of the circuit of Fig.~\ref{fig:U-bipartite-graph-search} associated with $\lambda_0$, we obtain a reduced circuit that provides a good approximation of $U(t)$. Using Eqs.~(\ref{eq:initial_condiction}) and~(\ref{eq:lambda-bipartite}), the asymptotic overlap between $\ket{\lambda_0}$ and $\ket{ \psi(0)}$ is
\begin{equation}
\left|\langle \lambda_0 | \psi(0) \rangle\right|^2 = \frac{1}{16 N^2}+O\left(\frac{1}{N^3}\right),
\end{equation}
which shows that asymptotically we can remove the term $\lambda_0|\lambda_0 \rangle\langle \lambda_0|$ from Eq.~(\ref{eq:H-bipartite-lambda}) because  the initial condition is $\ket{ \psi(0)}$.  Consistently, $\left|\langle \lambda^+ | \psi(0) \rangle\right|^2+\left|\langle \lambda^- | \psi(0) \rangle\right|^2\rightarrow 1$ in the asymptotic limit. The same behavior can be observed in many other graphs, in which the asymptotic time complexity of the search algorithm is determined by only two eigenvectors of the modified Hamiltonian.

\begin{figure}[!ht]
\includegraphics[scale=0.55]{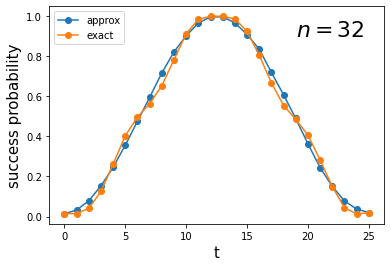}
\caption{Success probability $\left|\braket{0}{\psi(t)}\right|^2$ as a function of the number of steps $t$ for the search on $K_{n,n}$ with $n=32$. The orange curve is obtained using the exact decomposition~\eqref{eq:U-bipartite-graph-search} and the blue curve using the approximate circuit, both in Qiskit.}\label{fig:bipartite-n-8}
\end{figure}

Fig.~\ref{fig:bipartite-n-8} depicts the success probability $\left|\braket{0}{\psi(t)}\right|^2$ as a function of the number of steps $t$ for a complete bipartite graph  $K_{32,32}$. The orange curve is obtained using the implementation of the circuit of Fig.~\ref{fig:U-bipartite-graph-search} and the blue curve is obtained using the approximate circuit  (without the gates associated with $\lambda_0$), both using Qiskit. The curves are remarkably close even with $N=64$. The optimal running time is given by Eq.~(\ref{final_t_run}), which is
\begin{equation}
t_\text{opt}=\left\lfloor\frac{\pi}{2}\sqrt{N}\right\rfloor
\end{equation}
asymptotically. The blue curve is the square of a sinusoidal function, consistent with Eq.~\eqref{eq:p(t)}.

\section{Hypercube}\label{sec:hypercube}

The $n$-dimensional hypercube $Q_n$ is a graph whose vertex set is labeled by the binary $n$-tuples and two vertices are adjacent if their Hamming distance is exactly 1. The adjacency matrix is
\begin{equation}\label{eq:A-hypercube-graph}
   A =  \sum_{j=1}^{n} X_j,
\end{equation}
where $X_j$ is the Pauli-$X$ matrix acting on the $j$-th qubit.  The computational basis of the Hilbert space associated with $Q_n$ is spanned by the $n$-tuples and the dimension of the Hilbert space is  $N=2^n$. The evolution operator of the CTQW on $Q_n$ with no marked vertex at time $t$ is
\begin{equation}\label{eq:U-hypercube-graph-1}
    U(t) =  \big(R_{x}\left(-2\gamma t\right)\big)^{\otimes n},
\end{equation}
where $R_{x}(\theta) = \text{exp}(-i\theta X/2)$. The optimal value of $\gamma$ for the search algorithm on the hypercube is obtained in the next Subsection.



\subsection{Spatial search on hypercube}

To build the circuit of the spatial search algorithm on the hypercube, we use an approximate Hamiltonian (instead of~(\ref{eq:oracle_Hamiltonian})), which is given by
\begin{equation}\label{eq:H-hypercube-lambda}
{\mathcal{H}}_\text{approx}= \lambda^-|\lambda^- \rangle\langle \lambda^-|  +  \lambda^+|\lambda^+ \rangle\langle \lambda^+| ,
\end{equation}
where $\lambda^\pm$ are the eigenvalues of the first excited and ground states, respectively, of the exact Hamiltonian; $\ket{\lambda^\pm}$ are the associated normalized eigenvectors. We calculate a good approximation for $\lambda^\pm$ and $\ket{\lambda^\pm}$ by using the techniques of Appendix~\ref{appendix:B}. We disregard the remaining eigenvectors $\ket{\lambda}\not\in \text{span}\left\{\ket{\lambda^\pm}\right\}$ of the exact Hamiltonian because either they are associated with zero eigenvalues or $\left|\braket{\lambda}{\psi(0)}\right|\rightarrow 0$ asymptotically, where $\ket{\psi(0)}$ is the initial condition~\eqref{eq:initial_condiction}.

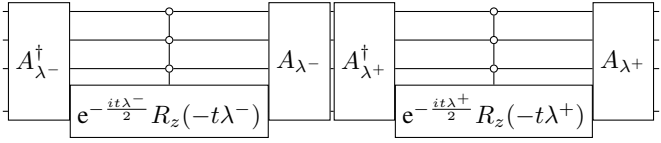
\begin{figure}[!ht]
\begin{tikzpicture}[scale=0.95]
\begin{yquant}
qubit {} q[4];
box {$A_{\lambda^-}^\dagger$} (q[0-3]) ;
box {$\text{e}^{-\frac{it\lambda^-}{2}}R_z(-t\lambda^-)$} q[3] ~ q[2-0];
box {$A_{\lambda^-}$} (q[0-3]) ;
box {$A_{\lambda^+}^\dagger$} (q[0-3]) ;
box {$\text{e}^{-\frac{it\lambda^+}{2}}R_z(-t\lambda^+)$} q[3] ~ q[2-0];
box {$A_{\lambda^+}$} (q[0-3]) ;
\end{yquant}
\end{tikzpicture}
\caption{\label{fig:U-hypercube-graph-search}The circuit that implements $U(t)$~(\ref{eq:U-hypercube-graph-search}) on the hypercube $Q_{n}$ with one marked vertex when $n=4$.  The circuit of $A_{\lambda^\pm}$ is depicted in Fig.\ref{fig:A-hypercube-graph};}
\end{figure}

The circuit that simulates the evolution operator
\begin{equation}\label{eq:U-hypercube-}
U(t)= \text{e}^{-i t{\mathcal{H}}_\text{approx}}
\end{equation}
is depicted in Fig.~\ref{fig:U-hypercube-graph-search}. To build this circuit, we replace~(\ref{eq:H-hypercube-lambda}) in (\ref{eq:U-hypercube-}) in order to obtain
\begin{equation}\label{eq:U-hypercube-graph-search}
U(t)=   A_{ \lambda^+}R_{\lambda^+}A_{\lambda^+}^\dagger A_{\lambda^-}R_{\lambda^-}A_{\lambda^-}^\dagger
\end{equation}
where
\begin{equation}
R_{\lambda^\pm} = I+(\text{e}^{-it\lambda^\pm}-1)|0\rangle\langle 0|
\end{equation}
and
\begin{equation}\label{eq:Alam-hypercube}
A_{\lambda^\pm}\ket{0}=|\lambda^\pm \rangle.
\end{equation}
The circuit of $A_{\lambda^\pm}$, which is depicted in Fig.~\ref{fig:A-hypercube-graph}, is obtained using the techniques described in Appendix~\ref{appendix:A} and $|\lambda^\pm \rangle$ using Appendix~\ref{appendix:B}.

\begin{figure}[!ht]
\begin{tikzpicture}[scale=1]
\begin{yquant*}
qubit {$\ket{0}$} q[4];
box {$H$} q[0-3];
box {$R_y(\theta^\pm_1)$} q[0];
box {$R_y(\theta^\pm_2)$} q[1] ~ q[0];
box {$R_y(\theta^\pm_3)$} q[2] ~ q[1-0];
box {$R_y(\theta^\pm_4)$} q[3] ~ q[2-0];
output {$\ket{\lambda^\pm}$} (q[0-3]);
\end{yquant*}
\end{tikzpicture}
\caption{\label{fig:A-hypercube-graph}The circuit that implements $A_{\lambda^\pm}$~(\ref{eq:Alam-hypercube}) when $n=4$. Angles $\theta^\pm_k$ are given by Eq.~(\ref{eq:thetap-hypercube}). }
\end{figure}
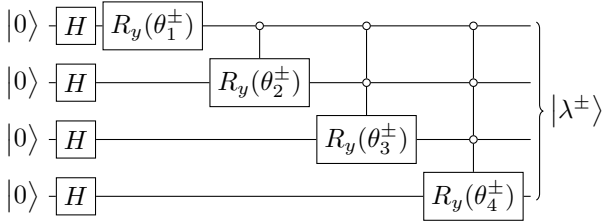

Since the characteristic polynomial of $A$~(\ref{eq:A-hypercube-graph}) is
\begin{equation}
\prod_{k=0}^n \left(\phi_k +2k-n\right)^{\binom{n}{k}},
\end{equation}
the eigenvalues are 
\begin{equation}
\phi_k = n-2k
\end{equation}
for $0\le k\le n$, with multiplicity $\binom{n}{k}$.
Since
\begin{equation}
\tilde A\ket{\ell}=(n-2|\ell|)\ket{\ell},
\end{equation}
where $\tilde A=H^{\otimes n}AH^{\otimes n}$ and $|\ell|$ is the Hamming weight of $\ell$, the eigenvectors associated with eigenvalue $(n-2k)$ are $H^{\otimes n}\ket{\ell}$ such that $|\ell|=k$. Define projector $P_k$ onto the $\phi_k$-eigenspace as
\begin{equation}
P_k= \sum_{\substack{\ell=0\\ |\ell|=k}}^{N-1}    H^{\otimes n}\ket{\ell}\bra{\ell}H^{\otimes n}.
\end{equation}
Using  these projectors, we obtain
\begin{align}
\left\| P_k\ket{w}   \right\|^2 &=\frac{1}{N^2}\sum_{y=0}^{N-1}\left|\sum_{|\ell|=k}(-1)^{\ell\cdot y}\right|^2
= \frac{1}{N}\binom{n}{k}.
\end{align}
Using formulas~(\ref{eq:S_1}) and~(\ref{eq:S_2}), we obtain
\begin{align}
S_1 &= \frac{1}{2N}\sum_{k=1}^n\frac{1}{k}\binom{n}{k}=\frac{1}{n}+\frac{1}{n^2}+O\left(\frac{1}{n^3}\right),\\
S_2 &= \frac{1}{4N}\sum_{k=1}^n\frac{1}{k^2}\binom{n}{k}=\frac{1}{n^2}+O\left(\frac{1}{n^3}\right).
\end{align}
The asymptotic expansion of $S_1$ is described in Ref.~\cite{BLP21} and of $S_2$ is obtained using the inequalities $\frac{1}{4N}\sum_{k=1}^n\frac{1}{(k+1)(k+2)}\binom{n}{k}\le S_2 \le S_1^2$.
Using~(\ref{eq:gamma}), we have $\gamma=S_1$ and then
\begin{equation}
\gamma = \frac{1}{n}+O\left(\frac{1}{n^2}\right).
\end{equation}
Using Eq.~(\ref{eq:epsilon}), we obtain
\begin{equation}
\epsilon = \frac{1}{\sqrt N}+O\left(\frac{1}{n\sqrt N}\right),
\end{equation}
and then using Eq.~(\ref{eq:lambdapm}), we obtain
\begin{equation}
\lambda^\pm = -1 \pm \frac{1}{\sqrt N}+O\left(\frac{1}{N}\right).
\end{equation}
Using Eq.~(\ref{eq:lambda_j_pm}), we obtain
\begin{equation}
\lambda^\pm(j) =\frac{1}{\sqrt 2}\left( {\delta_{jw}}  \mp \frac{1}{\sqrt{N}}\right).
\end{equation}

In the continuation, we take $w=0$.
Note that $\braket{\lambda^+}{\lambda^-}=0$ and $\braket{\lambda^\pm}{\lambda^\pm}=1\mp 2/\sqrt{N}$. The norm of theses eigenvectors tend to 1 asymptotically. 
Since $\ket{\lambda^\pm}$ belongs to  $\text{span}\left\{\ket{\alpha_0},...,\ket{\alpha_n}\right\}$ and minding the normalizing factor in Eq.~(\ref{eq:a-k-appen}), Eq.~\eqref{eq:ket-lambda} implies that ${\alpha_0}=\lambda^\pm(0)$ and ${\alpha_k}=\sqrt{2^{k-1}}\lambda^\pm(2^{k-1})$ ($1\le k\le n$).
Replacing ${\alpha_k}$ into Eq.~(\ref{eq:theta-k}), we obtain $\theta'^\pm_0=\pi/2$ and
\begin{equation}
\theta'^\pm_k = \mp \arctan \frac{1}{\sqrt{1+2^{1-k}N}}.
\end{equation}
for $1\le k\le n$.
Then, using Eq.~(\ref{eq:theta-k-prime}), we obtain
\begin{equation}\label{eq:thetap-hypercube}
\theta^\pm_k = \mp  2\arctan \frac{ 1}{\sqrt{1+2^k}} - \frac{\pi}{2}
\end{equation}
for $1\le k\le n$. Using $\theta^\pm_k$, we obtain the circuit for $A_{\lambda^\pm}$ such that $A_{\lambda^\pm}\ket{0} = \ket{\lambda^\pm}$,
which is depicted in Fig.~\ref{fig:A-hypercube-graph}.

\begin{figure}[!ht]
\includegraphics[scale=0.55]{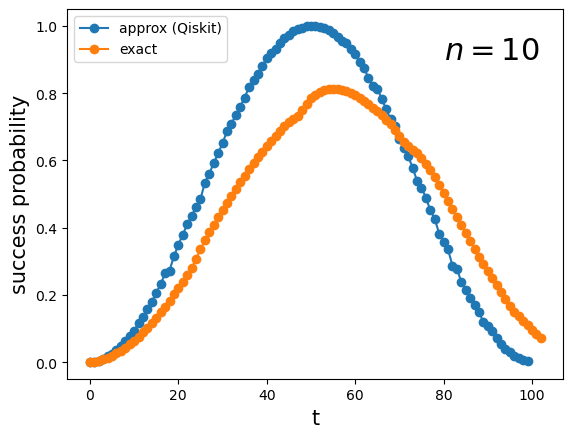}
\caption{Success probability $\left|\braket{0}{\psi(t)}\right|^2$ as a function of the number of steps $t$ for the search on $Q_n$ with $n=10$. The blue curve is obtained using the decomposition~\eqref{eq:U-hypercube-graph-search}, which is a good approximation of the success probability when using the exact Hamiltonian (orange curve).}\label{fig:hypercube-n-8}
\end{figure}

Fig.~\ref{fig:hypercube-n-8} depicts the success probability $\left|\braket{0}{\psi(t)}\right|^2$ as a function of the number of steps $t$ for a hypercube $Q_{10}$. The blue curve is obtained using the implementation of the circuit of Fig.~\ref{fig:U-hypercube-graph-search} in Qiskit and the orange curve is obtained using a numerical implementation of the exact Hamiltonian~(\ref{eq:oracle_Hamiltonian}).  The optimal running time is given by Eq.~(\ref{final_t_run}), which is
\begin{equation}
t_\text{opt}=\left\lfloor\frac{\pi}{2}\sqrt{N}\right\rfloor
\end{equation}
asymptotically.

\section{Final remarks}
\label{sec:conclusion}

In this work, we have described circuits that implement the evolution operator of the CTQW-based search algorithm on three graph classes: complete graphs, complete bipartite graphs, and hypercubes. We have implemented those circuits in Qiskit in order to compare with the analytic evolution operators. For the class of complete and complete bipartite graphs, the circuits implement the evolution operators exactly, modulo a global phase. For the class of hypercubes, the circuit implements an approximate circuit, so that the probability of finding the marked vertex is close to the probability obtained using the analytic evolution operator.

As a future work, we plan to go in three directions: (1)~we intend to apply our method to other graph classes, such as $N$-dimensional lattices, (2)~we intend to improve the depth of the circuits presented in this work, and (3)~we intend to implement those improved circuits on IBM quantum computers.

\appendix

\section{\label{appendix:A}State preparation}

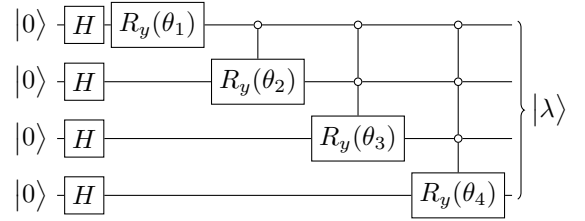
\begin{figure}[!ht]
\begin{tikzpicture}[scale=1.0]
\begin{yquant*}
qubit {$\ket{0}$} q[4];
box {$H$} q[0-3];
box {$R_y(\theta_1)$} q[0];
box {$R_y(\theta_2)$} q[1] ~ q[0];
box {$R_y(\theta_3)$} q[2] ~ q[1-0];
box {$R_y(\theta_4)$} q[3] ~ q[2-0];
output {$\ket{\lambda}$} (q[0-3]);
\end{yquant*}
\end{tikzpicture}
\caption{\label{fig:Ansatz}Ansatz for the circuit that implements an operator $A_{\lambda}$ with the property that $A_{\lambda}\ket{0}=\ket{\lambda}$  when $n=4$. Angles $\theta_k$ are obtained from Eqs.~(\ref{eq:theta-k-prime}) and~(\ref{eq:theta-k}). }
\end{figure}

There is a well-know recipe for obtaining a circuit of an operator $A$ that outputs a generic state $\ket{\lambda}$ of $n$ qubits when the input is $\ket{0}$, that is
\begin{equation}
A_{\lambda}\ket{0}=\ket{\lambda}.
\end{equation}
If $\ket{\lambda}$ has real entries and besides has many repeated entries, as discussed below, the Ansatz of Fig.~\ref{fig:Ansatz} provides a simpler circuit which has $O(n^2)$ basic gates assuming reasonable restrictions on $\theta_k$. The goal is to obtain angle $\theta_1,...,\theta_n$ that are used as parameters of the multi-controlled $R_y$ gates, where
\begin{equation}
R_y(\theta)=\text{e}^{-i \frac{\theta}{2} Y} =
    \begin{pmatrix}
        \cos\frac{\theta}{2}&-\sin\frac{\theta}{2}\vspace{3pt}\\
        \sin\frac{\theta}{2}&\cos\frac{\theta}{2}\\
    \end{pmatrix}.
\end{equation}

Define $\ket{\alpha_0}=\ket{0}$ and
\begin{equation}\label{eq:a-k-appen}
\ket{\alpha_k}=\frac{1}{\sqrt{2^{k-1}}} \sum_{j=2^{k-1}}^{2^k-1} \ket{j},
\end{equation}
for $1\le k\le n$, where $\{\ket{j}: 0\le j \le 2^n-1\}$ is the computational basis. If $\ket{\lambda}\in \text{span}\{\ket{\alpha_0},...,\ket{\alpha_n}\}$, then we can find $\theta_1,...,\theta_n$ such that $\ket{\lambda}$ is the output of the circuit of Fig.~\ref{fig:Ansatz}. To show this statement we use the fact that the output of the circuit is
\begin{equation}
\ket{\lambda} = \sum_{k=0}^n \sin \theta'_k \prod_{j=k+1}^n \cos \theta'_j \ket{\alpha_k},
\end{equation}
where
\begin{equation}\label{eq:theta-k-prime}
\theta_k=2\theta'_{n-k+1}-\frac{\pi}{2}
\end{equation}
for $1\le k \le n$ and $\theta'_0=\pi/2$. Assuming that 
\begin{equation}\label{eq:ket-lambda}
\ket{\lambda}= \sum_{k=0}^n{\alpha_k} \ket{\alpha_k},
\end{equation}
we obtain the system of equations
\begin{equation}\label{eq:alpha-k-append}
{\alpha_k}= \sin \theta'_k \prod_{j=k+1}^n \cos \theta'_j,
\end{equation}
for $0\le k\le n$ in terms of variables $\theta'_1,...,\theta'_n$. The solution is obtained in a recursive way starting with equation $\alpha_n=\sin \theta'_n$ and decreasing the index $k$ of Eq.~\eqref{eq:alpha-k-append}. We obtain
\begin{equation}
\sin\theta'_{k}=\frac{{\alpha_k}}{\sqrt{1-\sum_{j=k+1}^{n}{\alpha_j}^2}},
\end{equation}
and then
\begin{equation}\label{eq:theta-k}
\theta'_{k}=\arctan \frac{{\alpha_k}}{\sqrt{1-\sum_{j=k}^n{\alpha_j}^2}}.
\end{equation}

\section{Eigenvectors of the modified Hamiltonian}\label{appendix:B}

Suppose that the distinct eigenvalues of the adjacency matrix $A$ are $\phi_0>\phi_1>\dots>\phi_q$.
Define $P_{\ell}$ as the orthogonal projector onto the eigenspace of $A$ associated with eigenvalue $\phi_{\ell}$ for $0\leqslant \ell\leqslant q$, so that
\begin{equation}\label{eq:A-append-B}
	A = \sum_{\ell=0}^q \phi_{\ell} P_{\ell}.
\end{equation}

Let us assume that the asymptotic dynamics of the search algorithm depends only on two eigenvectors $\ket{\lambda^\pm}$ of $H$ associated with eigenvalues $\lambda^{\pm}$ closest of $-\gamma\phi_0$ so as
\begin{equation}\label{eq:lambdapm}
    \lambda^{\pm} = -\gamma\phi_0 \pm {\epsilon} + O\big(\epsilon^2\big),
\end{equation}
where $2\epsilon>0$ is  the spectral gap. Refs.~\cite{SPP22,LPST22} shows that the computational complexity of the search algorithm is determined by two sums given by
\begin{equation}\label{eq:S_1}
    S_1 = \sum_{\ell=1}^q \frac{\left\|P_{\ell}\ket{w}\right\|^2}{\phi_0-\phi_{\ell}}
\end{equation}
and 
\begin{equation}\label{eq:S_2}
    S_2 = \sum_{\ell=1}^q \frac{\left\|P_{\ell}\ket{w}\right\|^2}{(\phi_0-\phi_{\ell})^2},
\end{equation}
and the asymptotic value of $\gamma$ is obtained as
\begin{equation}\label{eq:gamma}
    \gamma = S_1
\end{equation}
and
\begin{equation}\label{eq:epsilon}
    \epsilon= \frac{S_1\left\|P_0\ket{w}\right\|}{\sqrt{S_2}}.
\end{equation}

Asymptotically, the success probability $\left|\braket{0}{\psi(t)}\right|^2$ is
\begin{equation}\label{eq:p(t)}
    p(t)=4\left|\braket{\lambda^+}{\psi(0)}\right|^2 \left|\braket{w}{\lambda^+}\right|^2\sin^2{\epsilon t},
\end{equation}
and the optimal running time is
\begin{equation}\label{final_t_run}
    t_\text{opt} = \frac{\pi}{2\epsilon}.
\end{equation}
To calculate the success probability, we need two extra quantities that are (1) the overlap between $\ket{\lambda^\pm}$ and the marked vertex, which is
\begin{equation}\label{eq:wlambda}
    \braket{w}{\lambda^\pm } = \frac{S_1}{\sqrt{2S_2}},
\end{equation}
and (2) the overlap between $\ket{\lambda^\pm}$ and the initial state, which is
\begin{equation}\label{eq:psi0lam}
     \braket{\psi(0)}{\lambda^\pm} = \mp\frac{1}{\sqrt{2N}  \left\| P_{0} \ket{w} \right\|}.
\end{equation}

Now we are ready to calculate an asymptotic expression for $\braket{j}{\lambda^\pm}$, that is, the $(j+1)$-th entry of $\ket{\lambda^\pm}$. These calculations go beyond the results of Refs.~\cite{SPP22,LPST22}.  Let us assume that the initial condition is a linear combination of $\ket{\lambda^+}$ and $\ket{\lambda^-}$ in the asymptotic limit. Using the completeness relation, we obtain
\begin{equation}\label{eq:jpsi0}
\braket{j}{\psi(0)} = \braket{\lambda^+}{\psi(0)}\,\lambda^+(j) +  \braket{\lambda^-}{\psi(0)}\,\lambda^-(j),
\end{equation}
where $\lambda^+(j)=\braket{j}{\lambda^+}$ and $\lambda^-(j)=\braket{j}{\lambda^-}$ are the unknowns. We obtain a second independent equation in these unknowns by sandwiching $H$, as follows
\begin{equation}\label{eq:jHpsi0}
\bra{j}H\ket{\psi(0)} = \lambda^+ \braket{\lambda^+}{\psi(0)}\,\lambda^+(j) +  \lambda^-\braket{\lambda^-}{\psi(0)}\,\lambda^-(j).
\end{equation}
Using Eq.~(\ref{eq:oracle_Hamiltonian}) and \eqref{eq:A-append-B}, we obtain
\begin{equation}
\bra{j}H\ket{\psi(0)} =- S- \frac{\delta_{jw}}{\sqrt{N}},
\end{equation}
where
\begin{equation}
S = \gamma\sum_\ell \phi_\ell \bra{j} P_\ell \ket{\psi(0)},
\end{equation}
and assuming that $\ket{\psi(0)}$ is the uniform state. The solution of the system of equations~(\ref{eq:jpsi0}) and~(\ref{eq:jHpsi0}) is
\begin{align}\label{eq:lambda_j_pm}
\lambda^\pm(j) &= \frac{\mp 1}{2\epsilon\sqrt{N} \braket{\lambda^\pm}{\psi(0)}}\left( \sqrt{N}S  +\delta_{jw}+\lambda^\mp\right).
\end{align}
The circuit of $A_{\lambda^\pm}$ such that $A_{\lambda^\pm}\ket{0}=\ket{\lambda^\pm}$ can be obtained using the techniques of Appendix~\ref{appendix:A}.

\section*{Acknowledgments}
We thank Pedro Lug\~ao and Frank Acasiete for interesting discussions. JKM acknowledges financial support from CNPq Grant PCI-DA number 304865/2019-2. RP acknowledges financial support from FAPERJ grant number CNE E-26/200.954/2021, and CNPq grant number 308923/2019-7.




\begin{thebibliography}{25}%
\makeatletter
\providecommand \@ifxundefined [1]{%
 \@ifx{#1\undefined}
}%
\providecommand \@ifnum [1]{%
 \ifnum #1\expandafter \@firstoftwo
 \else \expandafter \@secondoftwo
 \fi
}%
\providecommand \@ifx [1]{%
 \ifx #1\expandafter \@firstoftwo
 \else \expandafter \@secondoftwo
 \fi
}%
\providecommand \natexlab [1]{#1}%
\providecommand \enquote  [1]{``#1''}%
\providecommand \bibnamefont  [1]{#1}%
\providecommand \bibfnamefont [1]{#1}%
\providecommand \citenamefont [1]{#1}%
\providecommand \href@noop [0]{\@secondoftwo}%
\providecommand \href [0]{\begingroup \@sanitize@url \@href}%
\providecommand \@href[1]{\@@startlink{#1}\@@href}%
\providecommand \@@href[1]{\endgroup#1\@@endlink}%
\providecommand \@sanitize@url [0]{\catcode `\\12\catcode `\$12\catcode
  `\&12\catcode `\#12\catcode `\^12\catcode `\_12\catcode `\%12\relax}%
\providecommand \@@startlink[1]{}%
\providecommand \@@endlink[0]{}%
\providecommand \url  [0]{\begingroup\@sanitize@url \@url }%
\providecommand \@url [1]{\endgroup\@href {#1}{\urlprefix }}%
\providecommand \urlprefix  [0]{URL }%
\providecommand \Eprint [0]{\href }%
\providecommand \doibase [0]{https://doi.org/}%
\providecommand \selectlanguage [0]{\@gobble}%
\providecommand \bibinfo  [0]{\@secondoftwo}%
\providecommand \bibfield  [0]{\@secondoftwo}%
\providecommand \translation [1]{[#1]}%
\providecommand \BibitemOpen [0]{}%
\providecommand \bibitemStop [0]{}%
\providecommand \bibitemNoStop [0]{.\EOS\space}%
\providecommand \EOS [0]{\spacefactor3000\relax}%
\providecommand \BibitemShut  [1]{\csname bibitem#1\endcsname}%
\let\auto@bib@innerbib\@empty
\bibitem [{\citenamefont {Farhi}\ and\ \citenamefont {Gutmann}(1998)}]{FG98}%
  \BibitemOpen
  \bibfield  {author} {\bibinfo {author} {\bibfnamefont {E.}~\bibnamefont
  {Farhi}}\ and\ \bibinfo {author} {\bibfnamefont {S.}~\bibnamefont
  {Gutmann}},\ }\bibfield  {title} {\bibinfo {title} {Quantum computation and
  decision trees},\ }\href@noop {} {\bibfield  {journal} {\bibinfo  {journal}
  {Phys. Rev. A}\ }\textbf {\bibinfo {volume} {58}},\ \bibinfo {pages} {915}
  (\bibinfo {year} {1998})}\BibitemShut {NoStop}%
\bibitem [{\citenamefont {Childs}\ and\ \citenamefont
  {Goldstone}(2004)}]{CG_04}%
  \BibitemOpen
  \bibfield  {author} {\bibinfo {author} {\bibfnamefont {A.~M.}\ \bibnamefont
  {Childs}}\ and\ \bibinfo {author} {\bibfnamefont {J.}~\bibnamefont
  {Goldstone}},\ }\bibfield  {title} {\bibinfo {title} {Spatial search by
  quantum walk},\ }\href {https://doi.org/10.1103/PhysRevA.70.022314}
  {\bibfield  {journal} {\bibinfo  {journal} {Phys. Rev. A}\ }\textbf {\bibinfo
  {volume} {70}},\ \bibinfo {pages} {022314} (\bibinfo {year}
  {2004})}\BibitemShut {NoStop}%
\bibitem [{\citenamefont {Agliari}\ \emph {et~al.}(2010)\citenamefont
  {Agliari}, \citenamefont {Blumen},\ and\ \citenamefont {M\"ulken}}]{ABM_10}%
  \BibitemOpen
  \bibfield  {author} {\bibinfo {author} {\bibfnamefont {E.}~\bibnamefont
  {Agliari}}, \bibinfo {author} {\bibfnamefont {A.}~\bibnamefont {Blumen}},\
  and\ \bibinfo {author} {\bibfnamefont {O.}~\bibnamefont {M\"ulken}},\
  }\bibfield  {title} {\bibinfo {title} {Quantum-walk approach to searching on
  fractal structures},\ }\href {https://doi.org/10.1103/PhysRevA.82.012305}
  {\bibfield  {journal} {\bibinfo  {journal} {Phys. Rev. A}\ }\textbf {\bibinfo
  {volume} {82}},\ \bibinfo {pages} {012305} (\bibinfo {year}
  {2010})}\BibitemShut {NoStop}%
\bibitem [{\citenamefont {Philipp}\ \emph {et~al.}(2016)\citenamefont
  {Philipp}, \citenamefont {Tarrataca},\ and\ \citenamefont
  {Boettcher}}]{PTB_16}%
  \BibitemOpen
  \bibfield  {author} {\bibinfo {author} {\bibfnamefont {P.}~\bibnamefont
  {Philipp}}, \bibinfo {author} {\bibfnamefont {L.}~\bibnamefont {Tarrataca}},\
  and\ \bibinfo {author} {\bibfnamefont {S.}~\bibnamefont {Boettcher}},\
  }\bibfield  {title} {\bibinfo {title} {Continuous-time quantum search on
  balanced trees},\ }\href {https://doi.org/10.1103/PhysRevA.93.032305}
  {\bibfield  {journal} {\bibinfo  {journal} {Phys. Rev. A}\ }\textbf {\bibinfo
  {volume} {93}},\ \bibinfo {pages} {032305} (\bibinfo {year}
  {2016})}\BibitemShut {NoStop}%
\bibitem [{\citenamefont {Osada}\ \emph {et~al.}(2020)\citenamefont {Osada},
  \citenamefont {Coutinho}, \citenamefont {Omar}, \citenamefont {Sanaka},
  \citenamefont {Munro},\ and\ \citenamefont {Nemoto}}]{OCOSMN_20}%
  \BibitemOpen
  \bibfield  {author} {\bibinfo {author} {\bibfnamefont {T.}~\bibnamefont
  {Osada}}, \bibinfo {author} {\bibfnamefont {B.}~\bibnamefont {Coutinho}},
  \bibinfo {author} {\bibfnamefont {Y.}~\bibnamefont {Omar}}, \bibinfo {author}
  {\bibfnamefont {K.}~\bibnamefont {Sanaka}}, \bibinfo {author} {\bibfnamefont
  {W.~J.}\ \bibnamefont {Munro}},\ and\ \bibinfo {author} {\bibfnamefont
  {K.}~\bibnamefont {Nemoto}},\ }\bibfield  {title} {\bibinfo {title}
  {Continuous-time quantum-walk spatial search on the {B}ollob\'as scale-free
  network},\ }\href {https://doi.org/10.1103/PhysRevA.101.022310} {\bibfield
  {journal} {\bibinfo  {journal} {Phys. Rev. A}\ }\textbf {\bibinfo {volume}
  {101}},\ \bibinfo {pages} {022310} (\bibinfo {year} {2020})}\BibitemShut
  {NoStop}%
\bibitem [{\citenamefont {Farhi}\ \emph {et~al.}(2008)\citenamefont {Farhi},
  \citenamefont {Goldstone},\ and\ \citenamefont {Gutmann}}]{FGG_08}%
  \BibitemOpen
  \bibfield  {author} {\bibinfo {author} {\bibfnamefont {E.}~\bibnamefont
  {Farhi}}, \bibinfo {author} {\bibfnamefont {J.}~\bibnamefont {Goldstone}},\
  and\ \bibinfo {author} {\bibfnamefont {S.}~\bibnamefont {Gutmann}},\
  }\bibfield  {title} {\bibinfo {title} {A quantum algorithm for the
  {H}amiltonian {NAND} tree},\ }\href
  {https://doi.org/10.4086/toc.2008.v004a008} {\bibfield  {journal} {\bibinfo
  {journal} {Theory of Computing}\ }\textbf {\bibinfo {volume} {4}},\ \bibinfo
  {pages} {169} (\bibinfo {year} {2008})}\BibitemShut {NoStop}%
\bibitem [{\citenamefont {Dadras}\ \emph {et~al.}(2019)\citenamefont {Dadras},
  \citenamefont {Gresch}, \citenamefont {Groiseau}, \citenamefont {Wimberger},\
  and\ \citenamefont {Summy}}]{DGGCWS_19}%
  \BibitemOpen
  \bibfield  {author} {\bibinfo {author} {\bibfnamefont {S.}~\bibnamefont
  {Dadras}}, \bibinfo {author} {\bibfnamefont {A.}~\bibnamefont {Gresch}},
  \bibinfo {author} {\bibfnamefont {C.}~\bibnamefont {Groiseau}}, \bibinfo
  {author} {\bibfnamefont {S.}~\bibnamefont {Wimberger}},\ and\ \bibinfo
  {author} {\bibfnamefont {G.~S.}\ \bibnamefont {Summy}},\ }\bibfield  {title}
  {\bibinfo {title} {Experimental realization of a momentum-space quantum
  walk},\ }\href {https://doi.org/10.1103/PhysRevA.99.043617} {\bibfield
  {journal} {\bibinfo  {journal} {Phys. Rev. A}\ }\textbf {\bibinfo {volume}
  {99}},\ \bibinfo {pages} {043617} (\bibinfo {year} {2019})}\BibitemShut
  {NoStop}%
\bibitem [{\citenamefont {Delvecchio}\ \emph {et~al.}(2020)\citenamefont
  {Delvecchio}, \citenamefont {Groiseau}, \citenamefont {Petiziol},
  \citenamefont {Summy},\ and\ \citenamefont {Wimberger}}]{DGPSW20}%
  \BibitemOpen
  \bibfield  {author} {\bibinfo {author} {\bibfnamefont {M.}~\bibnamefont
  {Delvecchio}}, \bibinfo {author} {\bibfnamefont {C.}~\bibnamefont
  {Groiseau}}, \bibinfo {author} {\bibfnamefont {F.}~\bibnamefont {Petiziol}},
  \bibinfo {author} {\bibfnamefont {G.~S.}\ \bibnamefont {Summy}},\ and\
  \bibinfo {author} {\bibfnamefont {S.}~\bibnamefont {Wimberger}},\ }\bibfield
  {title} {\bibinfo {title} {Quantum search with a continuous-time quantum walk
  in momentum space},\ }\href {https://doi.org/10.1088/1361-6455/ab63ad}
  {\bibfield  {journal} {\bibinfo  {journal} {Journal of Physics B: Atomic,
  Molecular and Optical Physics}\ }\textbf {\bibinfo {volume} {53}},\ \bibinfo
  {pages} {065301} (\bibinfo {year} {2020})}\BibitemShut {NoStop}%
\bibitem [{\citenamefont {Wang}\ \emph {et~al.}(2020)\citenamefont {Wang},
  \citenamefont {Shi}, \citenamefont {Xiao}, \citenamefont {Wang},
  \citenamefont {Joglekar},\ and\ \citenamefont {Xue}}]{WSXWJX20}%
  \BibitemOpen
  \bibfield  {author} {\bibinfo {author} {\bibfnamefont {K.}~\bibnamefont
  {Wang}}, \bibinfo {author} {\bibfnamefont {Y.}~\bibnamefont {Shi}}, \bibinfo
  {author} {\bibfnamefont {L.}~\bibnamefont {Xiao}}, \bibinfo {author}
  {\bibfnamefont {J.}~\bibnamefont {Wang}}, \bibinfo {author} {\bibfnamefont
  {Y.~N.}\ \bibnamefont {Joglekar}},\ and\ \bibinfo {author} {\bibfnamefont
  {P.}~\bibnamefont {Xue}},\ }\bibfield  {title} {\bibinfo {title}
  {Experimental realization of continuous-time quantum walks on directed graphs
  and their application in pagerank},\ }\href
  {https://doi.org/10.1364/OPTICA.396228} {\bibfield  {journal} {\bibinfo
  {journal} {Optica}\ }\textbf {\bibinfo {volume} {7}},\ \bibinfo {pages}
  {1524} (\bibinfo {year} {2020})}\BibitemShut {NoStop}%
\bibitem [{\citenamefont {Benedetti}\ \emph {et~al.}(2021)\citenamefont
  {Benedetti}, \citenamefont {Tamascelli}, \citenamefont {Paris},\ and\
  \citenamefont {Crespi}}]{BTPC21}%
  \BibitemOpen
  \bibfield  {author} {\bibinfo {author} {\bibfnamefont {C.}~\bibnamefont
  {Benedetti}}, \bibinfo {author} {\bibfnamefont {D.}~\bibnamefont
  {Tamascelli}}, \bibinfo {author} {\bibfnamefont {M.~G.}\ \bibnamefont
  {Paris}},\ and\ \bibinfo {author} {\bibfnamefont {A.}~\bibnamefont
  {Crespi}},\ }\bibfield  {title} {\bibinfo {title} {Quantum spatial search in
  two-dimensional waveguide arrays},\ }\href
  {https://doi.org/10.1103/PhysRevApplied.16.054036} {\bibfield  {journal}
  {\bibinfo  {journal} {Phys. Rev. Applied}\ }\textbf {\bibinfo {volume}
  {16}},\ \bibinfo {pages} {054036} (\bibinfo {year} {2021})}\BibitemShut
  {NoStop}%
\bibitem [{\citenamefont {Qu}\ \emph {et~al.}(2022)\citenamefont {Qu},
  \citenamefont {Marsh}, \citenamefont {Wang}, \citenamefont {Xiao},
  \citenamefont {Wang},\ and\ \citenamefont {Xue}}]{QMWXWX22}%
  \BibitemOpen
  \bibfield  {author} {\bibinfo {author} {\bibfnamefont {D.}~\bibnamefont
  {Qu}}, \bibinfo {author} {\bibfnamefont {S.}~\bibnamefont {Marsh}}, \bibinfo
  {author} {\bibfnamefont {K.}~\bibnamefont {Wang}}, \bibinfo {author}
  {\bibfnamefont {L.}~\bibnamefont {Xiao}}, \bibinfo {author} {\bibfnamefont
  {J.}~\bibnamefont {Wang}},\ and\ \bibinfo {author} {\bibfnamefont
  {P.}~\bibnamefont {Xue}},\ }\bibfield  {title} {\bibinfo {title}
  {Deterministic search on star graphs via quantum walks},\ }\href
  {https://doi.org/10.1103/PhysRevLett.128.050501} {\bibfield  {journal}
  {\bibinfo  {journal} {Phys. Rev. Lett.}\ }\textbf {\bibinfo {volume} {128}},\
  \bibinfo {pages} {050501} (\bibinfo {year} {2022})}\BibitemShut {NoStop}%
\bibitem [{\citenamefont {Qiang}\ \emph {et~al.}(2016)\citenamefont {Qiang},
  \citenamefont {Loke}, \citenamefont {Montanaro}, \citenamefont
  {Aungskunsiri}, \citenamefont {Zhou}, \citenamefont {O'Brien}, \citenamefont
  {Wang},\ and\ \citenamefont {Matthews}}]{QLMAZOWM16}%
  \BibitemOpen
  \bibfield  {author} {\bibinfo {author} {\bibfnamefont {X.}~\bibnamefont
  {Qiang}}, \bibinfo {author} {\bibfnamefont {T.}~\bibnamefont {Loke}},
  \bibinfo {author} {\bibfnamefont {A.}~\bibnamefont {Montanaro}}, \bibinfo
  {author} {\bibfnamefont {K.}~\bibnamefont {Aungskunsiri}}, \bibinfo {author}
  {\bibfnamefont {X.}~\bibnamefont {Zhou}}, \bibinfo {author} {\bibfnamefont
  {J.~L.}\ \bibnamefont {O'Brien}}, \bibinfo {author} {\bibfnamefont {J.~B.}\
  \bibnamefont {Wang}},\ and\ \bibinfo {author} {\bibfnamefont {J.~C.~F.}\
  \bibnamefont {Matthews}},\ }\bibfield  {title} {\bibinfo {title} {Efficient
  quantum walk on a quantum processor},\ }\href
  {https://doi.org/10.1038/ncomms11511} {\bibfield  {journal} {\bibinfo
  {journal} {Nature Communications}\ }\textbf {\bibinfo {volume} {7}},\
  \bibinfo {pages} {11511} (\bibinfo {year} {2016})}\BibitemShut {NoStop}%
\bibitem [{\citenamefont {Santos}\ \emph {et~al.}(2021)\citenamefont {Santos},
  \citenamefont {Chagas},\ and\ \citenamefont {Chaves}}]{SCC21}%
  \BibitemOpen
  \bibfield  {author} {\bibinfo {author} {\bibfnamefont {J.}~\bibnamefont
  {Santos}}, \bibinfo {author} {\bibfnamefont {B.}~\bibnamefont {Chagas}},\
  and\ \bibinfo {author} {\bibfnamefont {R.}~\bibnamefont {Chaves}},\
  }\bibfield  {title} {\bibinfo {title} {Quantum walks in a superconducting
  quantum computer},\ }in\ \href {https://doi.org/10.5753/wquantum.2021.17223}
  {\emph {\bibinfo {booktitle} {Proceedings of WQUANTUM}}}\ (\bibinfo
  {publisher} {SBC},\ \bibinfo {address} {Porto Alegre, RS, Brasil},\ \bibinfo
  {year} {2021})\ pp.\ \bibinfo {pages} {25--30}\BibitemShut {NoStop}%
\bibitem [{\citenamefont {Aharonov}\ \emph {et~al.}(1993)\citenamefont
  {Aharonov}, \citenamefont {Davidovich},\ and\ \citenamefont
  {Zagury}}]{ADZ93}%
  \BibitemOpen
  \bibfield  {author} {\bibinfo {author} {\bibfnamefont {Y.}~\bibnamefont
  {Aharonov}}, \bibinfo {author} {\bibfnamefont {L.}~\bibnamefont
  {Davidovich}},\ and\ \bibinfo {author} {\bibfnamefont {N.}~\bibnamefont
  {Zagury}},\ }\bibfield  {title} {\bibinfo {title} {Quantum random walks},\
  }\href {https://doi.org/10.1103/PhysRevA.48.1687} {\bibfield  {journal}
  {\bibinfo  {journal} {Phys. Rev. A}\ }\textbf {\bibinfo {volume} {48}},\
  \bibinfo {pages} {1687} (\bibinfo {year} {1993})}\BibitemShut {NoStop}%
\bibitem [{\citenamefont {Portugal}(2016)}]{Por16b}%
  \BibitemOpen
  \bibfield  {author} {\bibinfo {author} {\bibfnamefont {R.}~\bibnamefont
  {Portugal}},\ }\bibfield  {title} {\bibinfo {title} {Staggered quantum walks
  on graphs},\ }\href {https://doi.org/10.1103/PhysRevA.93.062335} {\bibfield
  {journal} {\bibinfo  {journal} {Phys. Rev. A}\ }\textbf {\bibinfo {volume}
  {93}},\ \bibinfo {pages} {062335} (\bibinfo {year} {2016})}\BibitemShut
  {NoStop}%
\bibitem [{\citenamefont {Acasiete}\ \emph {et~al.}(2020)\citenamefont
  {Acasiete}, \citenamefont {Agostini}, \citenamefont {Moqadam},\ and\
  \citenamefont {Portugal}}]{AAMP20}%
  \BibitemOpen
  \bibfield  {author} {\bibinfo {author} {\bibfnamefont {F.}~\bibnamefont
  {Acasiete}}, \bibinfo {author} {\bibfnamefont {F.~P.}\ \bibnamefont
  {Agostini}}, \bibinfo {author} {\bibfnamefont {J.~K.}\ \bibnamefont
  {Moqadam}},\ and\ \bibinfo {author} {\bibfnamefont {R.}~\bibnamefont
  {Portugal}},\ }\bibfield  {title} {\bibinfo {title} {Implementation of
  quantum walks on {IBM} quantum computers},\ }\href
  {https://doi.org/10.1007/s11128-020-02938-5} {\bibfield  {journal} {\bibinfo
  {journal} {Quantum Information Processing}\ }\textbf {\bibinfo {volume}
  {19}},\ \bibinfo {pages} {426} (\bibinfo {year} {2020})}\BibitemShut
  {NoStop}%
\bibitem [{\citenamefont {Nzongani}\ \emph {et~al.}(2022)\citenamefont
  {Nzongani}, \citenamefont {Zylberman}, \citenamefont {Doncecchi},
  \citenamefont {Pérez}, \citenamefont {Debbasch},\ and\ \citenamefont
  {Arnault}}]{NZDPDA22}%
  \BibitemOpen
  \bibfield  {author} {\bibinfo {author} {\bibfnamefont {U.}~\bibnamefont
  {Nzongani}}, \bibinfo {author} {\bibfnamefont {J.}~\bibnamefont {Zylberman}},
  \bibinfo {author} {\bibfnamefont {C.-E.}\ \bibnamefont {Doncecchi}}, \bibinfo
  {author} {\bibfnamefont {A.}~\bibnamefont {Pérez}}, \bibinfo {author}
  {\bibfnamefont {F.}~\bibnamefont {Debbasch}},\ and\ \bibinfo {author}
  {\bibfnamefont {P.}~\bibnamefont {Arnault}},\ }\bibfield  {title} {\bibinfo
  {title} {Quantum circuits for discrete-time quantum walks with
  position-dependent coin operator}\ }\href
  {https://doi.org/10.48550/arxiv.2211.05271} {10.48550/arxiv.2211.05271}
  (\bibinfo {year} {2022})\BibitemShut {NoStop}%
\bibitem [{\citenamefont {Georgopoulos}\ \emph {et~al.}(2021)\citenamefont
  {Georgopoulos}, \citenamefont {Emary},\ and\ \citenamefont
  {Zuliani}}]{GEZ21}%
  \BibitemOpen
  \bibfield  {author} {\bibinfo {author} {\bibfnamefont {K.}~\bibnamefont
  {Georgopoulos}}, \bibinfo {author} {\bibfnamefont {C.}~\bibnamefont
  {Emary}},\ and\ \bibinfo {author} {\bibfnamefont {P.}~\bibnamefont
  {Zuliani}},\ }\bibfield  {title} {\bibinfo {title} {Comparison of
  quantum-walk implementations on noisy intermediate-scale quantum computers},\
  }\href {https://doi.org/10.1103/PhysRevA.103.022408} {\bibfield  {journal}
  {\bibinfo  {journal} {Phys. Rev. A}\ }\textbf {\bibinfo {volume} {103}},\
  \bibinfo {pages} {022408} (\bibinfo {year} {2021})}\BibitemShut {NoStop}%
\bibitem [{\citenamefont {Tanaka}\ \emph {et~al.}(2022)\citenamefont {Tanaka},
  \citenamefont {Sabri},\ and\ \citenamefont {Portugal}}]{TSP22}%
  \BibitemOpen
  \bibfield  {author} {\bibinfo {author} {\bibfnamefont {H.}~\bibnamefont
  {Tanaka}}, \bibinfo {author} {\bibfnamefont {M.}~\bibnamefont {Sabri}},\ and\
  \bibinfo {author} {\bibfnamefont {R.}~\bibnamefont {Portugal}},\ }\bibfield
  {title} {\bibinfo {title} {Spatial search on {J}ohnson graphs by
  continuous-time quantum walk},\ }\href
  {https://doi.org/10.1007/s11128-022-03417-9} {\bibfield  {journal} {\bibinfo
  {journal} {Quantum Information Processing}\ }\textbf {\bibinfo {volume}
  {21}},\ \bibinfo {pages} {74} (\bibinfo {year} {2022})}\BibitemShut {NoStop}%
\bibitem [{Note1()}]{Note1}%
  \BibitemOpen
  \bibinfo {note} {\protect \url {https://qiskit.org/}}\BibitemShut {NoStop}%
\bibitem [{Note2()}]{Note2}%
  \BibitemOpen
  \bibinfo {note} {See package \protect \textit
  {qiskit.circuit.QuantumCircuit}.}\BibitemShut {Stop}%
\bibitem [{Note3()}]{Note3}%
  \BibitemOpen
  \bibinfo {note} {The choice of $\protect \big | 0 \protect \big \rangle $ in
  this equation has no relation with the location of the marked
  vertex.}\BibitemShut {Stop}%
\bibitem [{\citenamefont {Bezerra}\ \emph {et~al.}(2021)\citenamefont
  {Bezerra}, \citenamefont {Lug\~ao},\ and\ \citenamefont {Portugal}}]{BLP21}%
  \BibitemOpen
  \bibfield  {author} {\bibinfo {author} {\bibfnamefont {G.~A.}\ \bibnamefont
  {Bezerra}}, \bibinfo {author} {\bibfnamefont {P.~H.~G.}\ \bibnamefont
  {Lug\~ao}},\ and\ \bibinfo {author} {\bibfnamefont {R.}~\bibnamefont
  {Portugal}},\ }\bibfield  {title} {\bibinfo {title} {Quantum-walk-based
  search algorithms with multiple marked vertices},\ }\href
  {https://doi.org/10.1103/PhysRevA.103.062202} {\bibfield  {journal} {\bibinfo
   {journal} {Phys. Rev. A}\ }\textbf {\bibinfo {volume} {103}},\ \bibinfo
  {pages} {062202} (\bibinfo {year} {2021})}\BibitemShut {NoStop}%
\bibitem [{\citenamefont {Silva}\ \emph {et~al.}(2022)\citenamefont {Silva},
  \citenamefont {Posner},\ and\ \citenamefont {Portugal}}]{SPP22}%
  \BibitemOpen
  \bibfield  {author} {\bibinfo {author} {\bibfnamefont {C.~F.~T.}\
  \bibnamefont {Silva}}, \bibinfo {author} {\bibfnamefont {D.}~\bibnamefont
  {Posner}},\ and\ \bibinfo {author} {\bibfnamefont {R.}~\bibnamefont
  {Portugal}},\ }\bibfield  {title} {\bibinfo {title} {Walking on vertices and
  edges by continuous-time quantum walk},\ }\href
  {https://arxiv.org/abs/2206.03375} {\bibfield  {journal} {\bibinfo  {journal}
  {10.48550/arxiv.2206.03375}\ } (\bibinfo {year} {2022})}\BibitemShut
  {NoStop}%
\bibitem [{\citenamefont {Lugão}\ \emph {et~al.}(2022)\citenamefont {Lugão},
  \citenamefont {Portugal}, \citenamefont {Sabri},\ and\ \citenamefont
  {Tanaka}}]{LPST22}%
  \BibitemOpen
  \bibfield  {author} {\bibinfo {author} {\bibfnamefont {P.~H.~G.}\
  \bibnamefont {Lugão}}, \bibinfo {author} {\bibfnamefont {R.}~\bibnamefont
  {Portugal}}, \bibinfo {author} {\bibfnamefont {M.}~\bibnamefont {Sabri}},\
  and\ \bibinfo {author} {\bibfnamefont {H.}~\bibnamefont {Tanaka}},\
  }\bibfield  {title} {\bibinfo {title} {Multimarked spatial search by
  continuous-time quantum walk}\ }\href
  {https://doi.org/10.48550/arxiv.2203.14384} {10.48550/arxiv.2203.14384}
  (\bibinfo {year} {2022})\BibitemShut {NoStop}%
\end{thebibliography}

%

\end{document}